# Phase Separation and Magnetic Order in K-doped Iron Selenide Superconductor

Wei Li[1], Hao Ding[1], Peng Deng[1], Kai Chang[1], Canli Song[1], Ke He[2], Lili Wang[2], Xucun Ma[2], Jiang-Ping Hu[3], Xi Chen[1, *], and Qi-Kun Xue[1, *]

[1]*State Key Laboratory of Low-Dimensional Quantum Physics, Department of Physics, Tsinghua University, Beijing 100084, China*
[2]*Institute of Physics, Chinese Academy of Sciences, Beijing 100190, China*
[3]*Department of Physics, Purdue University, West Lafayette, Indiana 47907, USA*

Alkali-doped iron selenide is the latest member of high $T_c$ superconductor family, and its peculiar characters have immediately attracted extensive attention. We prepared high-quality potassium-doped iron selenide ($K_xFe_{2-y}Se_2$) thin films by molecular beam epitaxy and unambiguously demonstrated the existence of phase separation, which is currently under debate, in this material using scanning tunneling microscopy and spectroscopy. The stoichiometric superconducting phase $KFe_2Se_2$ contains no iron vacancies, while the insulating phase has a √5×√5 vacancy order. The iron vacancies are shown always destructive to superconductivity in $KFe_2Se_2$. Our study on the subgap bound states induced by the iron vacancies further reveals a magnetically-related bipartite order in the superconducting phase. These findings not only solve the existing controversies in the atomic and electronic structures in $K_xFe_{2-y}Se_2$, but also provide valuable information on understanding the superconductivity and its interplay with magnetism in iron-based superconductors.

*\* To whom correspondence should be addressed. Email: xc@mail.tsinghua.edu.cn, qkxue@mail.tsinghua.edu.cn*

The newly discovered alkali-doped iron selenide superconductors[1,2] not only reach a superconducting transition temperature as high as 32 K, but also exhibit several unique characters that are noticeably absent in other iron-based superconductors, such as antiferromagnetically ordered insulating phases[3,4], extremely high Neel transition temperatures[5,6], and the presence of intrinsic Fe vacancies and ordering[7-9]. These features have generated considerable excitements as well as confusions, regarding the delicate interplay between Fe vacancies, magnetism and superconductivity[10-12], on which there is no consensus to date. Here we report on molecular beam epitaxy (MBE) growth of high-quality $K_xFe_{2-y}Se_2$ thin films and *in situ* low-temperature scanning tunneling microscope (STM) measurement of their atomic and electronic structures. We unambiguously demonstrate that a $K_xFe_{2-y}Se_2$ sample contains two distinct phases: an insulating phase with well-defined $\sqrt{5}\times\sqrt{5}$ order of Fe vacancies, and a superconducting $KFe_2Se_2$ phase containing no Fe vacancies. We further show that the vacancy and magnetism are strongly coupled and an individual Fe vacancy can locally destroy superconductivity in a similar way as a magnetic impurity in conventional superconductors. The measurement of magnetic field dependence of the Fe-vacancy-induced bound states explicitly reveals a hidden magnetically-related bipartite order in the tetragonal iron lattice. These findings elucidate the existing controversies on this new superconductor and provide atomistic information on understanding the interplay between magnetism and superconductivity in iron-based superconductors.

Our experiments were conducted in a Unisoku ultra-high vacuum (UHV) low temperature (down to 0.4 K) STM equipped with an MBE chamber for thin film growth. Magnetic field up to 11 T can be applied perpendicular to the sample surface. We grew $K_xFe_{2-y}Se_2$ thin films on the graphitized 6H-SiC(0001) substrate. High purity Fe (99.995%), Se (99.9999%) and K were evaporated from two standard

Knudsen cells and one SAES getter, respectively. The growth was carried out under Se-rich condition with a nominal Se/Fe flux ratio of ~20, which is known to lead to stoichiometric and superconductive FeSe[13]. To intercalate K atoms in between FeSe layers (see the schematic crystal structure of $KFe_2Se_2$ in Fig. 1A), the substrate was held at 440°C during growth and the sample was subsequently annealed at 470°C for three hours. The K flux is relatively flexible, which determines the area ratio of two different phases as shown below. The STM image in Fig. 1B shows the typical atomically flat surface of a sample.

Two distinct regions (marked by I and II in Fig. 1B), coexisting side by side, are clearly revealed in the film, indicating that phase separation[7, 12] occurs. The STM image with atomic resolution (Fig. 1C) of region I exhibits a centered rectangular lattice structure. The periods along the two orthogonal directions are 5.5 Å and 14.1 Å, respectively, as shown in Fig. 1C. Comparison with X-ray diffraction data suggests that the orientation of the film is (110) (see Fig. 1D) instead of the natural cleavage plane (001). Thus the STM actually images the cross-section of the layered material. The (110) plane is terminated with K and Se atoms. The K atoms are visible at positive bias and form atomic rows (Fig. 1C), which are 7.05 Å apart and oriented along the $[1\bar{1}0]$ direction. At negative bias, the Se atomic rows (indicated by black dots in Fig. 1E) appear and zigzag through the K atoms. These images are fully consistent with the atomic structure of the (110) surface (Fig. 1D). Furthermore, our STM observation shows that there are very few defects in region I. We therefore identify region I as the stoichiometric $KFe_2Se_2$.

The scanning tunneling spectroscopy (STS) probes the quasiparticle density of states (DOS). In Fig. 1F, we show the STS at 0.4 K in region I. The spectrum exhibits a superconducting gap centered at the Fermi energy and two characteristic coherence peaks, indicating that the stoichiometric phase $KFe_2Se_2$ is a superconductor. As expected, the gap and coherence peaks gradually disappear at higher temperatures as shown in the temperature-dependent tunneling spectra (fig. S1). Similar to cuprate[14] and iron-pnictide[15,16] superconductors, the electron-hole asymmetry is clearly noticeable in STS.

The STS reveals a double-gap structure. The larger gap $\Delta_1$=4 meV is half of the energy between the two pronounced coherence peaks. The smaller one, roughly $\Delta_2$~1 meV, is estimated by the two shoulders near the Fermi energy in the spectra. The coherence peaks of the smaller gap are hardly distinguishable because of thermal broadening. Nearly isotropic gaps have also been observed by angle-resolved photoemission spectroscopy (ARPES)[17-20] on the electron-like Fermi surfaces near M and Γ points, giving rise to $\Delta_1$, whereas $\Delta_2$ is too small to be resolved by ARPES.

The finite-size effect on superconductivity is negligible because the typical size of region I, 20 nm thick (along the [110] direction in *a-b* plane) and 50 nm wide (along the *c* and [$1\bar{1}0$] directions), is much larger than the superconducting coherence length of $KFe_2Se_2$. Here the anisotropic coherence lengths $\xi_{ab}$~2 nm and $\xi_c$~0.5 nm can be estimated by the upper critical fields[21,22] in the *a-b* plane and along the *c* direction using $H_{c2}^{ab}=\Phi_0/2\pi\xi_{ab}\xi_c$ and $H_{c2}^{c}=\Phi_0/2\pi\xi_{ab}^2$, respectively. On the other hand, the penetration depth[23] of the material is much larger than the sample size. Under this circumstance, the magnetic field is essentially uniform in a sample. Different from our previous work on FeSe[13], no vortex core has been resolved in the zero-bias conductance map of $KFe_2Se_2$ probably due to the large penetration depth. However, the effect of magnetic field on superconductivity in $KFe_2Se_2$ is still clearly manifested in STS. Magnetic field breaks time-reversal symmetry and weakens superconductivity, giving rise to a reduction in the height of coherence peak in DOS (see fig. S2).

Because of the good sample quality, the superconducting gap exhibits high spatial homogeneity, as shown in Figs. 1G and 1H. Previous studies of iron-based superconductors usually suffer from various imperfections in the materials[24]. It is therefore very crucial to prepare high-quality samples. As mentioned above, we solved this problem by MBE growth under well-controlled conditions (see also Ref. 13). In addition, surface contamination is avoided since our film growth and STM study were conducted in a single UHV system. More significantly, by growing films with (110)-oriented crystallographic surface where both K and Se atoms are exposed, we have realized the cross-sectional tunneling configuration. Such configuration for unconventional superconductors has been highly desirable, but very challenging to achieve due to the difficulty in preparing a suitable surface for STM studies[25].

Now we turn to region II, which shows periodic stripe pattern (Fig. 1B). The STS in Fig. 2A exhibits an energy gap up to 0.43 eV across the Fermi level, suggesting that this region is insulating. Besides the K atomic rows in the topmost layer, there is a superposed striped structure with a period of 14.0 Å in the STM image (Fig. 2B). The stripes are along the c-axis and perpendicular to the K atomic rows. We attribute this superstructure to the $\sqrt{5}\times\sqrt{5}$ pattern of Fe vacancies in the second atomic layer. They are visible in STM images because the electronic structure of the topmost layer is perturbed by the missing Fe-Se bonds. Neutron scattering[5], transmission electron microscopy[7] and X-ray diffraction[8] measurements have all revealed a $\sqrt{5}\times\sqrt{5}$ order formed by Fe vacancies. Such a blocked checkerboard pattern[9] gives rise to a $5a_{Fe}\sim14.0$ Å periodicity in the (110) plane (Fig. 2C) and is consistent with the superstructure observed in our STM images. Here $a_{Fe}$ is the in-plane distance between two neighboring Fe atoms. The $\sqrt{5}\times\sqrt{5}$ vacancy order leads to a composition of $K_xFe_{1.6}Se_2$, where $x$ is either 1 or 0.8. As illustrated in fig. S3, among every five consecutive bright spots along the K atomic row, the one in the middle appears different from the other four. The STM image alone cannot tell if this difference is due to the missing of one K atom ($x$=0.8) or simply electronic feature ($x$=1), thus leaving the value of $x$ undetermined.

The above STM study, which combines the capabilities of imaging and spectroscopy together, has thus explicitly proved that $K_xFe_{2-y}Se_2$ is composed of two different phases, i.e., superconducting $KFe_2Se_2$ and insulating $K_xFe_{1.6}Se_2$. The $\sqrt{5}\times\sqrt{5}$ ordered pattern of Fe vacancies only exists in the insulating phase. As discussed below, our further experiment demonstrates that the superconducting phase has a hidden order related to Fe as well.

Defect-induced subgap states in superconductors, which can be detected by local probes such as STM/STS[26-28], have often been used to uncover the nature of superconducting state and magnetic interaction. To reveal the hidden order, we introduced defects (bright parallelograms in Fig. 3A) into the superconducting $KFe_2Se_2$ phase by annealing the sample at 450°C for several hours. The atomically resolved STM images indicate that the topmost layer remains perfect lattice without K or Se vacancies (fig. S4). The resulted defects are always located in the middle between two adjacent K atomic rows. We therefore attribute the parallelogram-shaped structures to single Fe vacancies in the second atomic layer (Fig. 3B). By examining the registration of Fe sites with respect to the Se lattice in the topmost layer, the Fe atoms in the (110) plane can be divided into two interpenetrating sublattices. The Fe vacancies on two different sublattices are labeled by A and B, respectively. The atomic structure, chemical environment and STM images of these two types of vacancies are mirror-images of each other.

A Fe vacancy carries spin and breaks superconducting pairing in the singlet channel through spin-flip scattering. The exchange interaction $JS \cdot S_{vac}$, where $S$ and $S_{vac}$ are the spins of a quasiparticle and a vacancy, respectively, gives rise to the bound quasiparticle states. The STS on a vacancy (Fig. 3C) shows strongly suppressed coherence peaks and a pair of resonances inside the superconducting gap, i.e., an electron-like bound state at 1.9 mV and a hole-like bound state at -1.9 mV. While the energies of the electron-like and hole-like states are symmetric with respect to zero bias, their amplitudes are different as a result of on-site Coulomb interaction[29]. The spatial extent of a resonance peak as shown in the *dI/dV* mapping (Fig. 3D) is comparable with the coherence lengths in $[1\bar{1}0]$ and [001] directions. A vacancy is considered to be isolated if no others exist within twice the coherence length. The spectra on all isolated vacancies are identical, indicating that type-A and B vacancies have the same magnitude of magnetic moment and the same exchange interaction *J*. The spectrum itself does not provide direct information on the orientation of magnetic moment.

To reveal the spin orientations of the two types (A and B) of vacancies, we apply a magnetic field perpendicular to the sample surface to break the rotational symmetry. To date, the magnetic field effect on defect-induced subgap resonance has not been observed. However, in $KFe_2Se_2$, the observation of such effect becomes possible owing to the high upper critical field and large penetration depth. As clearly shown in Figs. 3E and 3F, the energy of the subgap resonance in $KFe_2Se_2$ exhibits a linear dependence on magnetic field ***B***: $E=E_0+g\mu_B \bm{B} \cdot \bm{S}$, where $\mu_B$ is the Bohr magneton and *g* the Landé factor. Fitting the data to a line (Fig. 3G) gives $g=2.07\pm0.07$, close to 2.0023 of a free electron. The most striking behavior of the field effect is that the peaks on type-A and B vacancies shift to opposite directions with magnetic field. The opposite shifting suggests that the two types of vacancies have different spin orientations, implying a magnetically-related bipartite order in the tetragonal Fe lattice. Such bipartite structure may account for the large electron-like Fermi surface sheet with weak intensity near Γ point in ARPES data[19,20] through Brillouin zone folding.

The interplay between magnetism and superconductivity is one of the most intriguing phenomena in high $T_c$ superconductors. The technique mentioned above can sensitively probe local magnetism in superconductors and may help to understand how superconductivity arises in high $T_c$ materials.

We have demonstrated that a single Fe vacancy locally suppresses Cooper pairing. The role of Fe vacancies in the superconducting phase has been rather controversial in previous studies. For example, some experiment suggested that randomly distributed vacancies may help to stabilize the superconducting state[12]. To further elucidate the effect of Fe vacancies on superconductivity, we prepared samples with higher density of randomly distributed vacancies (see Fig. 4A) by UHV annealing. The STS in Fig. 4B shows that a sample with high density of vacancies eventually becomes a gapless superconductor. Therefore the Fe vacancies are always destructive to superconductivity in $KFe_2Se_2$.

**Acknowledgements:** We thank Y. G. Zhao for helpful discussions. The work was financially supported by National Science Foundation and Ministry of Science and Technology of China.

**Figure Captions**

Fig. 1. MBE film and the superconducting phase of $K_xFe_{2-y}Se_2$. (A) The crystal structure of $KFe_2Se_2$. The same conventions for atoms and Miller indices are used throughout. (B) STM topographic image (bias voltage V=3.9 V, tunneling current $I_t$=0.02 nA) of a $K_xFe_{2-y}Se_2$ film. The grain size is usually larger than 50 nm × 50 nm. Two distinct regions are labeled by I and II, respectively. (C) Atomic-resolution STM topography of region I (5 nm × 5 nm, V=0.15 V, $I_t$=0.03 nA). The K atoms are visible at positive bias. (D) Atomic structure of (110) plane. K and Se atoms are in the topmost layer. Fe atoms are in the second layer. (E) STM image of the same area as (C), but at negative bias. V=-2 V, $I_t$=-0.03 nA. Se atoms (marked by black dots) are visible. The positions of K atoms are marked by white dots. (F) Differential conductance spectrum in region I measured at 0.4 K (setpoint: V=14 mV, $I_t$=0.1 nA). (G-H) Uniformity of the superconducting gap. The spectra in (H) were measured along the white line in (G). The size of image in (G) is 20 nm × 20 nm. The setpoint: V=14 mV, $I_t$=0.1 nA.

Fig. 2. The insulating phase. (A) Differential conductance spectrum in region II (setpoint: V=1.1 V, $I_t$=0.26 nA). (B) Atomic-resolution STM topography of region II (10 nm × 10 nm, V=2.9 V, $I_t$=0.02 nA). (C) The structure of $\sqrt{5}\times\sqrt{5}$ Fe vacancy pattern as seen from (001) and (110) planes, respectively. The positions of Fe vacancies are marked by crosses.

Fig. 3. Fe vacancy-induced bound states in the superconducting gap of $KFe_2Se_2$. (A, B) STM topography (12 nm × 12 nm, V=30 mV, $I_t$=0.03 nA) and atomic structure of Fe vacancies. Two types of vacancies are labeled by A and B, respectively. (C, D) $dI/dV$ spectrum measured at 0.4 K (setpoint, V=15 mV, $I_t$=0.1 nA) and density of states map

of the bound state on a single Fe vacancy (4 nm × 4 nm, 2 mV). The tunneling junction was set by 25 mV and 0.16 nA during mapping. The positions of Fe vacancy, K atoms and Se atoms are marked a cross, black dots and white dots, respectively. (E to G) Magnetic field dependence of the bound state energies.

Fig. 4. Suppression of superconductivity of $KFe_2Se_2$ by Fe vacancies. (A) STM topography (18 nm × 18 nm, V=60 mV, $I_t$=0.02 nA) of an area with high density of Fe vacancies. (B) Superconducting gaps (setpoint: V=15 mV, $I_t$=0.1 nA) at various density of vacancies. Black curve: no vacancy. Cyan curve: low density ($0.05/nm^2$) of vacancies. Red curve: high density ($0.1/nm^2$) of vacancies as shown in (A).

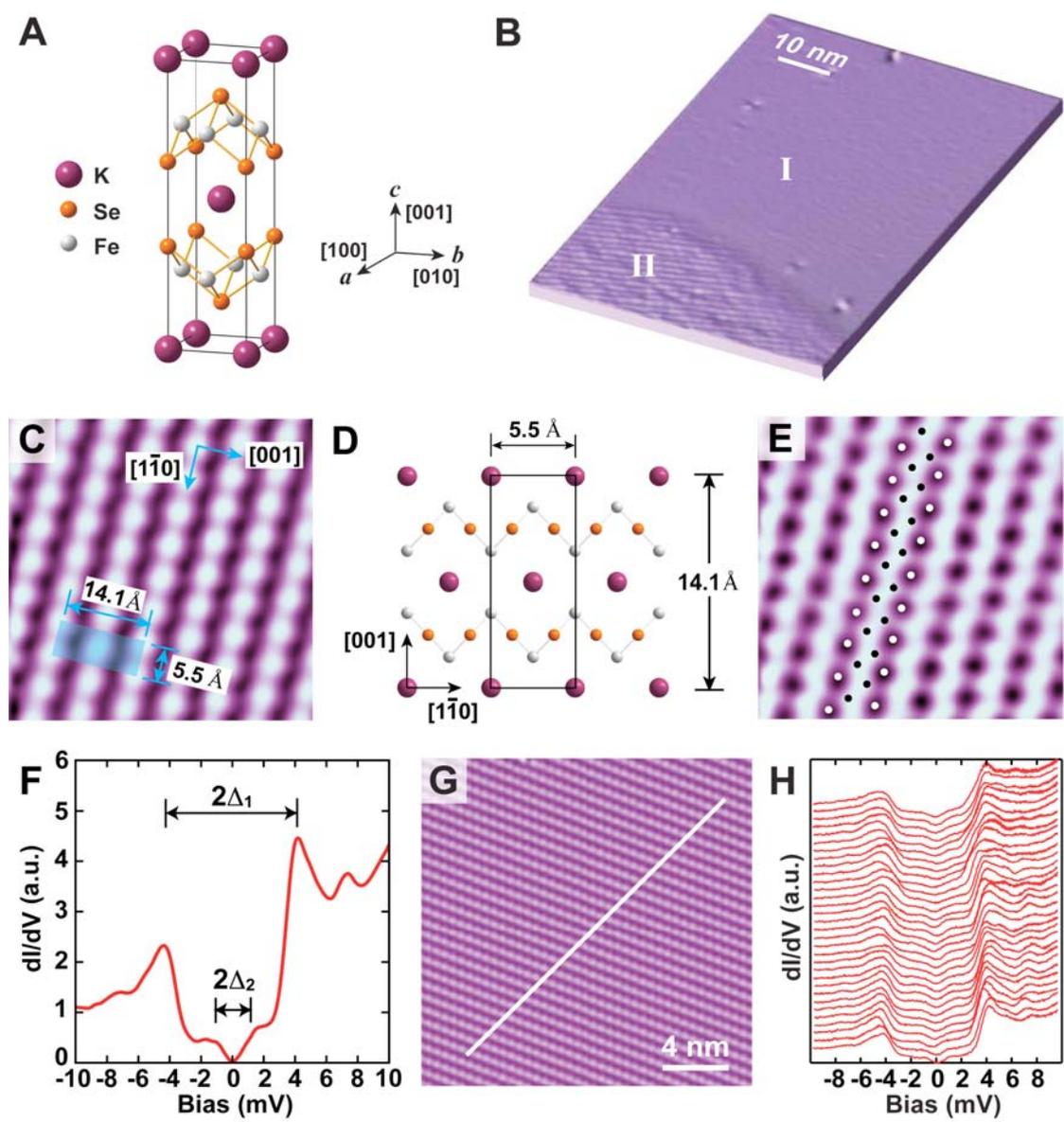

**Figure 1**

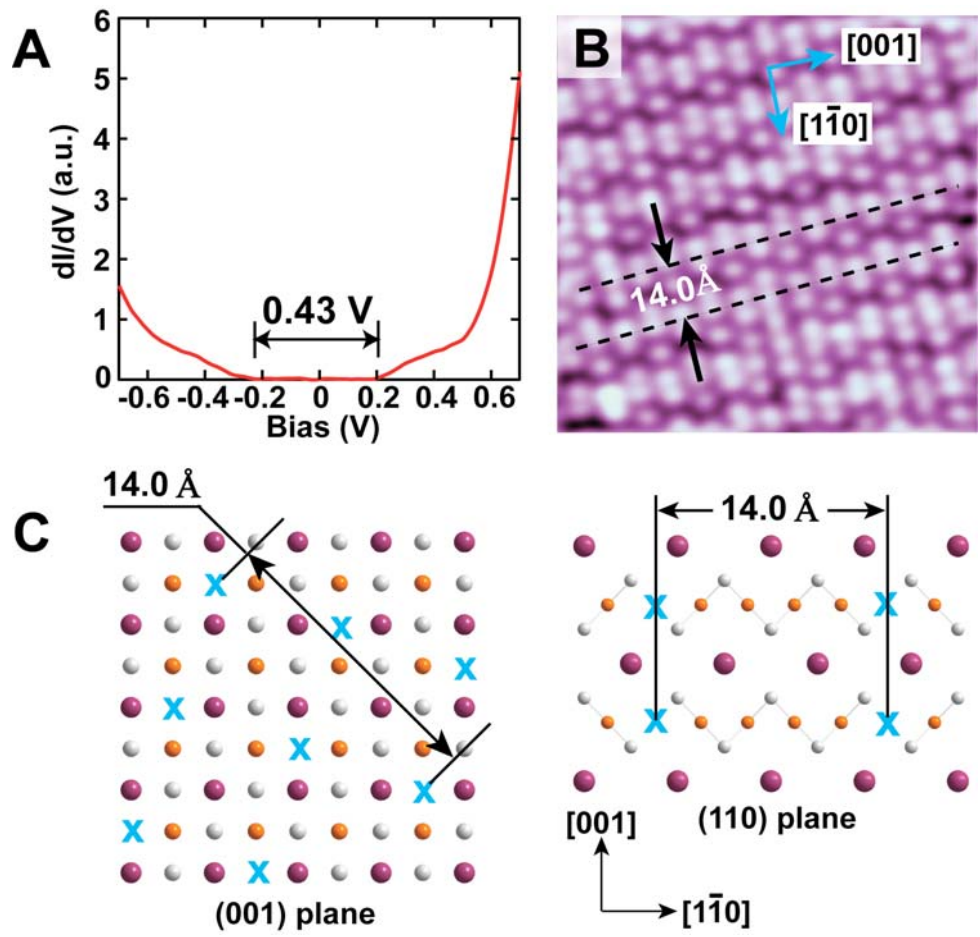

**Figure 2**

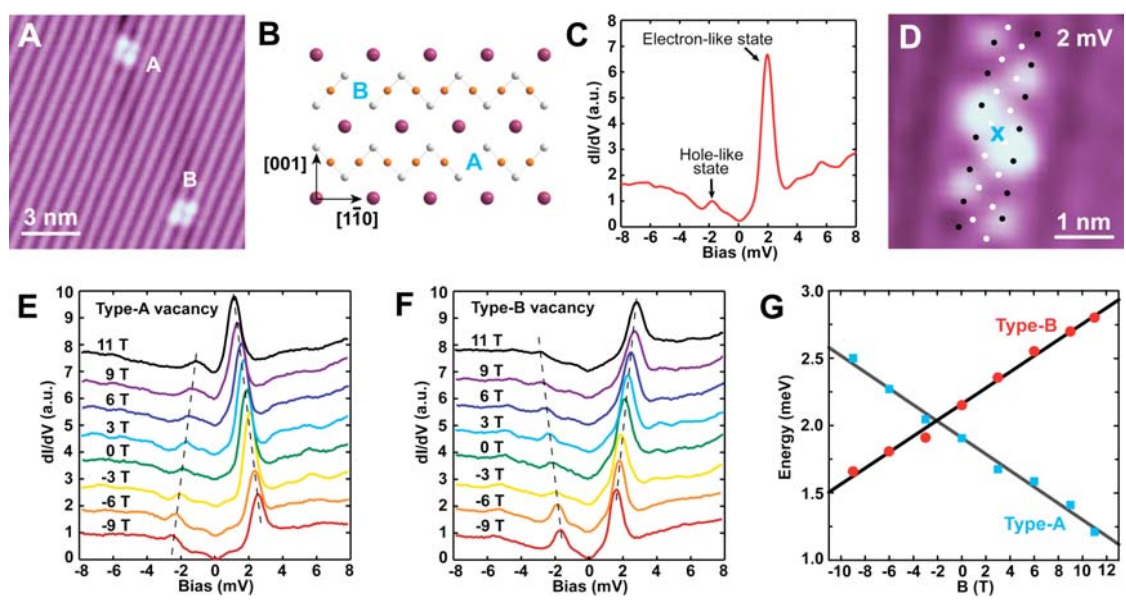

**Figure 3**

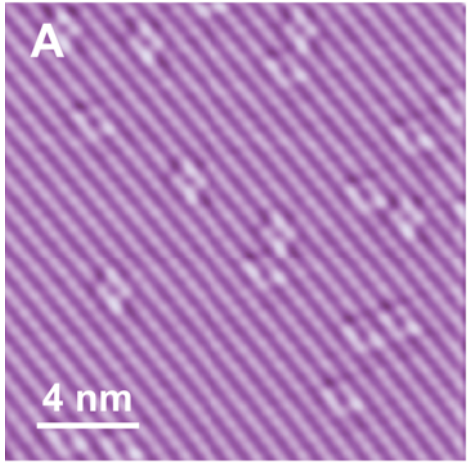 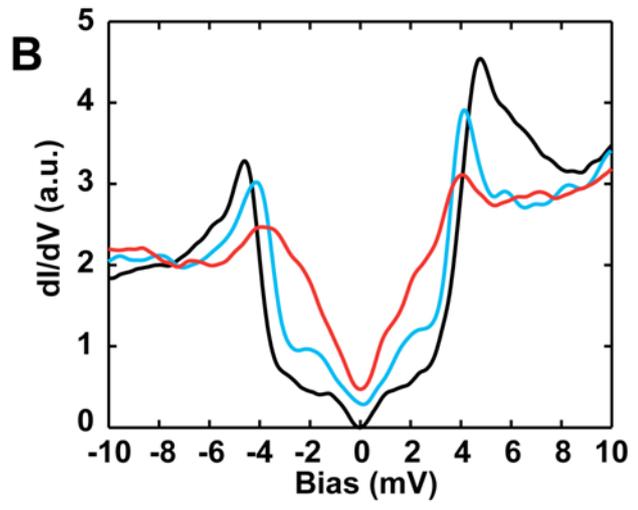

**Figure 4**

Supporting Materials for

# Phase Separation and Magnetic Order in K-doped Iron Selenide Superconductor

Wei Li[1], Hao Ding[1], Peng Deng[1], Kai Chang[1], Canli Song[1], Ke He[2], Lili Wang[2], Xucun Ma[2], Jiang-Ping Hu[3], Xi Chen[1, *], and Qi-Kun Xue[1, *]

[1]*State Key Laboratory of Low-Dimensional Quantum Physics, Department of Physics, Tsinghua University, Beijing 100084, China*
[2]*Institute of Physics, Chinese Academy of Sciences, Beijing 100190, China*
[3]*Department of Physics, Purdue University, West Lafayette, Indiana 47907, USA*

*\* To whom correspondence should be addressed. Email: xc@mail.tsinghua.edu.cn, qkxue@mail.tsinghua.edu.cn*

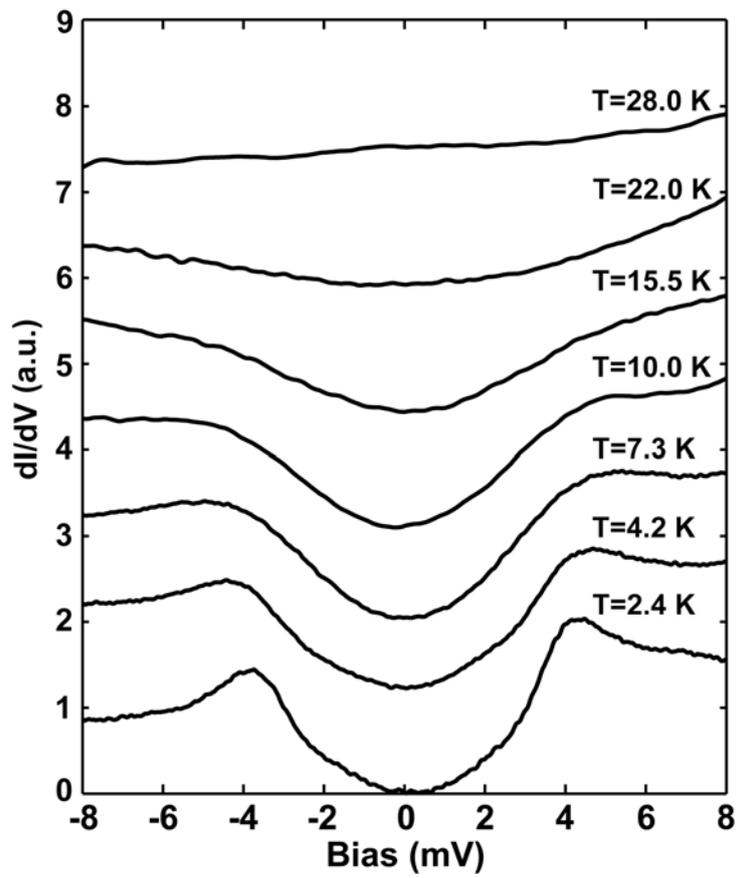

**Fig. S1.** Temperature dependence of differential tunneling conductance of the superconducting phase KFe$_2$Se$_2$. Setpoint: V=25 mV, I$_t$=0.16 nA. The curves are offset vertically for clarity.

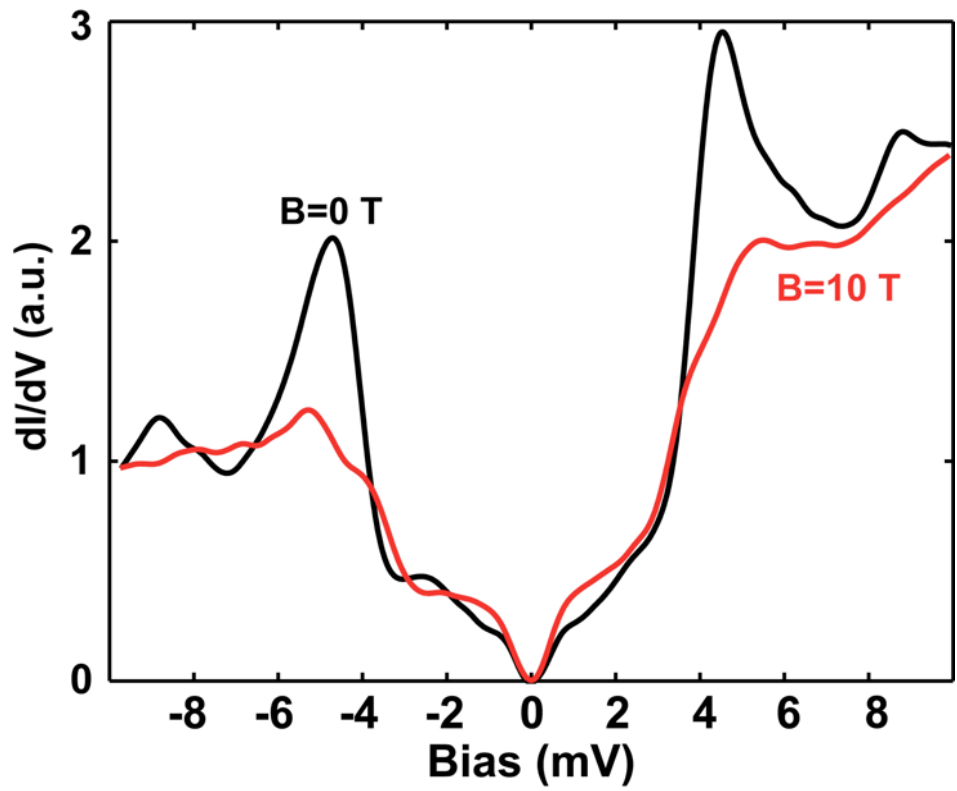

**Fig. S2.** Suppression of coherence peaks of the superconducting phase $KFe_2Se_2$ by magnetic field. Setpoint: V=14 mV, $I_t$=0.1 nA.

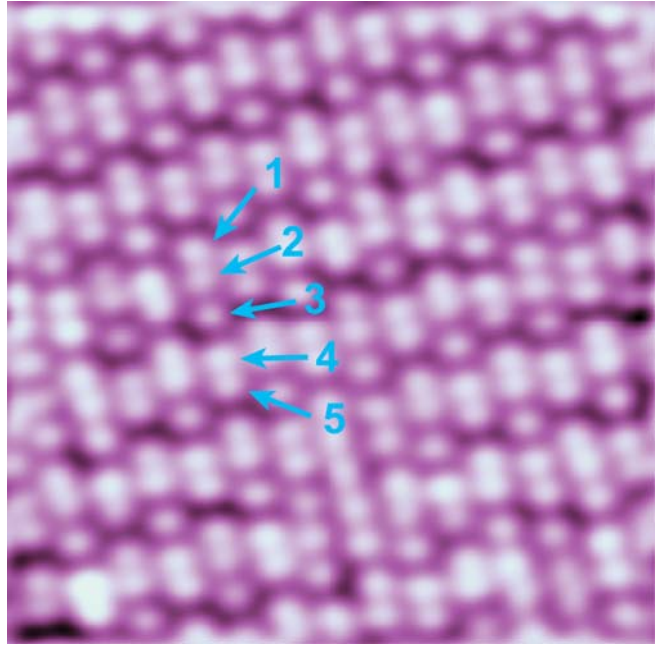

**Fig. S3.** STM topography of the insulating phase $K_xFe_{1.6}Se_2$ (x=1 or 0.8). The bright spot marked by "3" is different from the other 4 spots for K atoms.

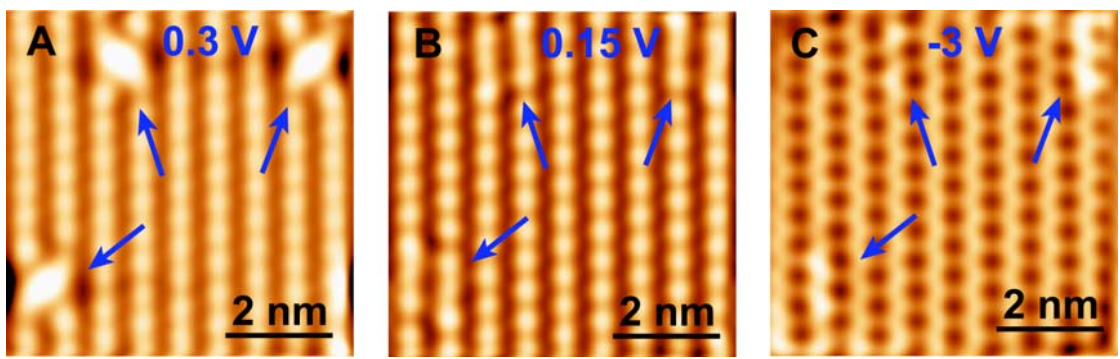

**Fig. S4.** Atomic structure of the topmost layer above Fe vacancies (indicated by blue arrows). (A) STM topographic image of an area with vacancies. (B) STM image of the same area at V=0.15 V showing the continuity of K atomic chains. (C) STM image of the same area at V=-3 V showing the continuity of Se atomic chains.